\documentclass[reprint,superscriptaddress,amsmath,amssymb,aps,prd,notitlepage,longbibliography,floatfix,nofootinbib,onecolumn]{revtex4-1}

\usepackage{tensor}     
\usepackage{graphicx}   
\usepackage{subfigure}
\usepackage[
colorlinks=true,        
citecolor=blue,         
linkcolor=blue,         
urlcolor=blue           
]{hyperref}             
\usepackage{bm}         
\usepackage{xcolor}     
\usepackage{lipsum}
\usepackage{color}      
\usepackage[utf8]{inputenc} 
\usepackage[section]{placeins} 
\usepackage{multirow}
\usepackage[title]{appendix}
\newcommand{\nc}{\newcommand*}

\nc{\al}{\alpha}
\nc{\s}{\sigma}
\nc{\kp}{\kappa}
\nc{\dt}{\delta}
\nc{\Dt}{\Delta}
\nc{\Ld}{\Lambda}
\nc{\p}{\partial}
\nc{\Gm}{\Gamma}
\nc{\om}{\omega}
\nc{\Om}{\Omega}
\nc{\rd}{\mathrm{d}}
\def\({\left(}
\def\){\right)}
\def\[{\left[}
\def\]{\right]}
\def\e{\begin{equation}}
\def\q{\end{equation}}
\def\m{\begin{eqnarray}}
\def\n{\end{eqnarray}}
\nc{\Eq}[1]{Eq.~\eqref{#1}}     
\nc{\Fig}[1]{Fig.~\ref{#1}}     
\nc{\Table}[1]{Table~\ref{#1}}  
\nc{\Sec}[1]{Sec.~\ref{#1}}     
\nc{\Msun}{M_\odot}             
\nc{\fpbh}{f_{\mathrm{pbh}}}    
\nc{\mpbh}{m_{\mathrm{pbh}}}    
\nc{\fpbhn}{f_{\mathrm{pbh0}}}    
\nc{\mR}{\mathcal{R}} 
\nc{\seq}{\sigma_{\mathrm{eq}}}
\nc{\ogw}{\Omega_{\mathrm{GW}}}
\nc{\gpcyr}{\mathrm{Gpc}^{-3}\,\mathrm{yr}^{-1}}
\nc{\lvc}{LIGO/Virgo} 
\nc{\SNR}{\mathrm{SNR}} 
\nc{\mmin}{{m_{\mathrm{min}}}}
\nc{\mmax}{{m_{\mathrm{max}}}}
\nc{\Mmin}{{M_{\mathrm{min}}}}
\nc{\fmin}{{f_{\mathrm{min}}}}
\nc{\VT}{\mathrm{VT}}
\nc{\rhoGW}{\rho_{\mathrm{GW}}}
\nc{\vth}{\vec{\theta}}
\nc{\vd}{\vec{d}}
\nc{\vla}{\vec{\lambda}}
\nc{\Nobs}{N_{\mathrm{obs}}}
\nc{\av}[1]{\langle #1 \rangle} 
\nc{\km}{\mathrm{km}}
\nc{\Mpc}{\mathrm{Mpc}}
\nc{\Tobs}{T_{\mathrm{obs}}}
\nc{\Ntemp}{N_{\mathrm{temp}}}
\nc{\fyr}{f_{\mathrm{yr}}}
\nc{\addref}{[\textcolor{red}{add ref}] } 
\nc{\eg}{\textit{e.g.~}}
\nc{\app}{\approx}
\nc{\hf}{\frac{1}{2}}
\nc{\discuss}{\textcolor{red}{Add discussion here!}}
\nc{\red}[1]{\textcolor{red}{#1}}
\nc{\dd}{\mathrm{d}}
\nc{\Tr}{\text{Tr}}
\nc{\hp}{h_+} 
\nc{\hc}{h_{\times}} 
\nc{\Oh}{\hat{\Omega}}
\nc{\vx}{\vec{x}}
\nc{\mh}{\hat{m}}
\nc{\nh}{\hat{n}}
\nc{\zh}{\hat{z}}
\nc{\ph}{\hat{p}}
\nc{\A}[1]{\mathcal{A}_{#1}}
\nc{\Ogw}[1]{\Omega_{\mathrm{#1}}}
\nc{\bn}[1]{\dt\bm{t}_{\text{#1}}}
\nc{\bC}[1]{\bm{C}_{\text{#1}}}
\nc{\NTOA}{N_{\text{TOA}}}
\nc{\Nmode}{{N_{\text{mode}}}}
\nc{\ARN}{A_{\rm{RN}}}
\nc{\gRN}{\gamma_{\rm{RN}}}
\nc{\bS}{\mathbf{\Sigma}}
\nc{\br}{\mathbf{r}}
\nc{\bN}{\mathbf{R}}
\nc{\Agw}{A_\mathrm{GWB}}
\nc{\UCP}{\mathrm{UCP}}
\nc{\TT}{\mathrm{TT}}
\nc{\ST}{\mathrm{ST}}
\nc{\SL}{\mathrm{SL}}
\nc{\VL}{\mathrm{VL}}
\nc{\mH}{\mathcal{H}}
\nc{\BFST}{$107 \pm 7$}

\begin{document}
	
\title{{One-Loop Tensor Power Spectrum from a Non-Canonical Spectator Field during Inflation}}


\author{Zhe Li}
\email{lizhe232@mails.ucas.ac.cn}
\affiliation{School of Fundamental Physics and Mathematical Sciences
Hangzhou Institute for Advanced Study, UCAS, Hangzhou 310024, China}
\affiliation{School of Physical Sciences,
	University of Chinese Academy of Sciences,
	No. 19A Yuquan Road, Beijing 100049, China}
\affiliation{CAS Key Laboratory of Theoretical Physics,
Institute of Theoretical Physics, Chinese Academy of Sciences,
Beijing 100190, China}
\author{Chen Yuan}
\email{Corresponding author: chenyuan@tecnico.ulisboa.pt}
\affiliation{CENTRA, Departamento de Física, Instituto Superior Técnico – IST, Universidade de Lisboa – UL, Avenida Rovisco Pais 1, 1049–001 Lisboa, Portugal}

\author{Qing-Guo Huang}
\email{Corresponding author: huangqg@itp.ac.cn}
\affiliation{School of Fundamental Physics and Mathematical Sciences
Hangzhou Institute for Advanced Study, UCAS, Hangzhou 310024, China}
\affiliation{School of Physical Sciences,
University of Chinese Academy of Sciences,
No. 19A Yuquan Road, Beijing 100049, China}
\affiliation{CAS Key Laboratory of Theoretical Physics,
Institute of Theoretical Physics, Chinese Academy of Sciences,
Beijing 100190, China}

\date{\today}

\begin{abstract}
We compute the full one-loop corrections to the primordial tensor power spectrum in an inflationary scenario with a {non-canonical spectator field}, using the in-in formalism. We derive semi-analytic results for the scalar-sourced one-loop tensor spectrum and the effective tensor-to-scalar ratio, $r_{\mathrm{eff}}$. We consider two representative coupling functions: a localized Gaussian dip (Model G), which leads to moderate loop corrections, and a rapidly oscillatory coupling (Model O), which can yield much larger loop contributions. For Model G, we find a $\mathcal{O}(1)$ correction to $r_{\mathrm{eff}}$ while Model O can significantly enhance $r_{\mathrm{eff}}$ by several orders of magnitude (relative to the tree-level value).
We further calculate the energy density of primordial gravitational waves. Assuming that primordial black holes with mass $10^{-12}M_{\odot}$ generated in this scenario, constitute all of the dark matter, we find that the results are several orders of magnitude lower than the sensitivities of Taiji/TianQin/LISA.
\end{abstract}

\maketitle

\maketitle

\section{Introduction}
The era of gravitational wave (GW) astronomy is inaugurated by the detection of GW events from binary black holes (BHs) and binary neutron star coalescences \cite{LIGOScientific:2016emj}. Since then, the LIGO-Virgo-KAGRA collaboration has reported a catalog of nearly a hundred GW events, opening a new window to explore the strong field regime of gravity \cite{LIGOScientific:2017vwq,LIGOScientific:2016sjg,LIGOScientific:2018mvr,LIGOScientific:2020ibl,KAGRA:2021vkt,LIGOScientific:2025slb}.

The idea of primordial black hole (PBH) has aroused interest as contributors to dark matter and a possible explanation for the origin of mass components for these GW events.
A standard way to form PBHs is through the gravitational collapse of overdense regions in the very early universe \cite{Zeldovich:1967lct,Hawking:1971ei,Carr:1974nx}. These overdensed regions originate from large curvature perturbations and will collapse to form PBHs immediately after the corresponding wavelength re-enters the horizon. To generate sufficient PBHs to explain all or a main fraction of the dark matter, these curvature perturbations are required to be enhanced by several orders of magnitude on small scales compared to those observed in the Cosmic Microwave Background (CMB) (see e.g., \cite{Sasaki:2018dmp,Green:2020jor} for reviews of PBHs).

Various inflation models have been proposed to generate PBHs which amplify the curvature power spectrum on small scales while maintaining the CMB consistency on large scales. In recent years, special attention has focused on single field inflation with an ultra-slow-roll (USR) phase followed by a transition back to slow roll (SR) (for details of USR inflation, see e.g., \cite{Leach:2001zf,Tsamis:2003px,Kinney:2005vj,Dimopoulos:2017ged}).
A recent line of work was ignited by \cite{Kristiano:2022maq}, the authors argued that a sharp USR $\!\to$ SR transition can cause a one-loop correction to the large-scale power comparable to the tree level, apparently challenging the perturbativity and PBH scenarios~\cite{Kristiano:2022maq}. 
Subsequent analyses revisited the one-loop power spectrum, showing that the correction is sensitive to the sharpness of the USR $\to$ SR transition. For instance, smooth and finite-time transitions can reduce the one-loop spectrum \cite{Firouzjahi:2023aum,Firouzjahi:2023bkt}. In parallel, the one-loop bispectrum was computed and found to have a local shape with an amplitude controlled by the same parameters (the USR $\to$ SR transition), becoming large only in the artificially sharp limit~\cite{Firouzjahi:2024psd}. 
A comprehensive in--in calculation including all relevant cubic/quartic interactions together with the counterterms confirms that the loop-to-tree ratio is governed by these time scales, and that in realistically smoothed transitions compatible with PBH production perturbation theory need not break down~\cite{Ballesteros:2024zdp}.

Beyond single field inflation, multi-field scenarios offer new possibilities. Among them, the {non-canonical spectator field} model stands out as a robust mechanism \cite{Cai:2021wzd,Lalak:2007vi,vandeBruck:2014ata,Braglia:2020fms,Pi:2021dft,Meng:2022low,Chen:2023lou,Bian:2025ifp,Pi:2017gih}. In this scenario, a spectator couples to the inflaton through a specific coupling function $f(\phi)$ which sources the enhancement of the spectator field to generate PBHs.
Although such a multi-field inflation can produce small non-Gaussianities \cite{Meng:2022low} and is expected to maintain the validity of perturbation theory \cite{Meng:2022ixx}, a significant enhancement of scalar fluctuations implies strong non-linear interactions between the scalar and tensor perturbations \cite{Inomata:2022yte}.
The quantum corrections of the primordial tensor spectrum due to the scalar perturbations have received far less attention. 
Although recent work has considered full quantum one-loop corrections to the tensor power spectrum~\cite{Ota:2022xni,Ota:2022hvh}, it relied on a generic excited scalar state without specifying a physical mechanism responsible for the small-scale enhancement. In contrast, our work is based on a concrete physical model, the {non-canonical spectator field} scenario, where the enhancement is naturally realized by the feature in the coupling function.

In this work we compute, from first principles, the one-loop tensor power spectrum in such {non-canonical spectator field} models, adopting a renormalization scheme based on Bunch--Davies (BD) subtraction with UV alignment to remove the UV tail of the mode functions. 
We analyze two representative coupling profiles: a single localized ``Gaussian dip'' (Model G) and a finite-time oscillating coupling (Model O).
Our goal is to investigate whether the one-loop primordial tensor spectrum under such scenario is still in consistent with current observations and, at the same time, generate sufficient PBHs to account for a main fraction of the dark matter.

This paper is organized as follows. We begin in \Sec{secII} by introducing the spectator field model and the cosmological setup. In \Sec{secIII} we detail the calculation of the one-loop tensor power spectrum using the in-in formalism. Our numerical results are presented in \Sec{secIV}, where we also discuss the renormalization. Finally, we summarize in \Sec{secV}.

\section{Cosmological Setup}
\label{secII}
In this paper, we consider a massive {non-canonical spectator field} model, which serves as a source for generating PBHs \cite{Pi:2021dft,Meng:2022low}. The action involving the inflaton field $\phi$ and the spectator field $\chi$ is given by
\begin{equation}
	S[\phi,\chi]=\int \mathrm d^4 x \sqrt{-g}\left[ -\frac{1}{2} g^{\mu\nu}\partial_\mu\phi\partial_\nu\phi-V(\phi)-\frac{1}{2}f^2(\phi)g^{\mu\nu}\partial_\mu\chi\partial_\nu\chi-\frac{1}{2}m^2\chi^2\right], 
	\label{equ:actionpc}
\end{equation}
where $f(\phi)$ denotes the {non-canonical} coupling function between the inflaton and the spectator. Follow the proposal in \cite{Pi:2021dft}, we consider a coupling $f(\phi)$ characterized by a narrow feature around $\phi=\phi_*$ and $f(\phi)=1$ out of the feature. Then the power spectrum of $\delta\chi$ can be significantly enhanced at small scales compared to the CMB scales. 
To illustrate the effects of $f(\phi)$ around $\phi_*$, we consider two phenomenological forms, one with Gaussian dip (Model G) proposed in \cite{Pi:2021dft} and the other one with oscillating feature (Model O) proposed in \cite{Meng:2022low}:
\m
f_\text{G}(\phi)=1-A_\text{G} \exp\[{-\frac{(\phi-\phi_*)^2}{2\Delta_\phi^2}}\],
\n
\m
f_\text{O}(\phi)=1-{A_\text{O}\over 2}\[\text{tanh}{\phi-(\phi_*-\Delta_\phi/2)\over \Lambda_\phi}-\text{tanh}{\phi-(\phi_*+\Delta_\phi/2)\over \Lambda_\phi}\]\sin {\phi-\phi_*\over \xi_\phi}, 
\n
Here, $A_\text{G}$ and $A_\text{O}$ control the amplitude of the feature. The evolution of $\phi$ around $\phi_*$ is approximately given by $\phi(\tau)\simeq \phi_*+\phi_*'(\tau-\tau_*)$, where $\phi_*'$ is the velocity of $\phi$ at the conformal time $\tau_*$ when $\phi=\phi_*$.

We calculate the tensor power spectrum in the uniform curvature gauge. The spatial metric is  written as
\begin{equation}
	\gamma_{ij} = a^2 \mathrm{e}^{h_{ij}} = a^2 \left[\delta_{ij}+h_{ij}+\frac{1}{2}h_i^{~k}h_{kj} + \mathcal{O}(h^3) \right]
\end{equation}
where $a(\tau)$ is the scale factor and $h_{ij}$ represents the transverse-traceless tensor perturbations, satisfying $\partial^ih_{ij}=0$ and $h^i_i=0$. 
{Due to the transverse-traceless nature of tensor perturbations, the determinant of the spatial metric remains unperturbed at non-linear orders, i.e.,
\begin{equation}
\det(e^{h_{ij}})=1.
\end{equation}
Therefore, the mass term does not generate direct tensor--scalar vertices in the uniform-curvature gauge. Indeed, the mass term is simply
\begin{equation}
S_{m}=-\frac{1}{2}\int d\tau d^3x\, a^4 m^2 \chi^2 .
\end{equation}
Expanding $\chi(\tau,\mathbf{x})=\chi(\tau)+\delta\chi(\tau,\mathbf{x})$, one finds that $S_m$ does not contain
$h\,\delta\chi\,\delta\chi$ or $hh\,\delta\chi\,\delta\chi$ interaction vertices. Therefore, the tensor--scalar interaction vertices relevant for the one-loop tensor spectrum arise entirely from the kinetic sector of $\chi$,
\begin{equation}\label{equ:actionint}
S_{\rm full}\supset -\frac{1}{2}\int d^4x\,\sqrt{-g}\, f^2(\phi) g^{\mu\nu}\partial_\mu\chi\partial_\nu\chi .
\end{equation}}
We denote the spectator scalar field fluctuation by $\delta\chi$. Applying Legendre transformation to \Eq{equ:actionint}, we arrive at the interaction Hamiltonian:
\m
H_{\mathrm{int}} \equiv H^{(3)}_{\mathrm{int}} + H^{(4)}_{\mathrm{int}} = a^{2}f^{2} \int \mathrm{d}^{3} x \left(-\frac{1}{2} h^{i j}+\frac{1}{4} h^{i k} h_{k}^{j}\right) \partial_{i} \delta \chi \partial_{j} \delta \chi. \label{equ:hamint}
\n

In Fourier space, the scalar and tensor perturbations can be written as:
\m
\begin{aligned}
\delta \chi(\tau, \mathbf{x}) =& \int \frac{\mathrm{d}^{3} q}{(2 \pi)^{3}} e^{i \mathbf{q} \cdot \mathbf{x}} \delta \chi_{\mathbf{q}}(\tau),
\end{aligned}
\n
\m
\begin{aligned}
h_{i j}(\tau, \mathbf{x}) =& \int \frac{\mathrm{d}^{3} q}{(2 \pi)^{3}} e^{i \mathbf{q} \cdot \mathbf{x}} \sum_{s=+,\times} e_{i j}^{s}(\hat{q}) h_{\mathbf{q}}^{s}(\tau),
\end{aligned}
\n
where $\hat{q} \equiv \mathbf{q} /|\mathbf{q}|$ and the polarization tensor $ e_{i j}^{s}(\hat{q})$ satisfies the orthogonality and completeness conditions. Substituting these expansions into the interaction action, we derive the third and fourth-order interaction Hamiltonians:
\m
\begin{aligned}
H_{\mathrm{int}}^{(3)}=&\frac{1}{2} \prod_{A=1}^{3}\left(\int \frac{\mathrm{d}^{3} p_{A}}{(2 \pi)^{3}}\right)(2 \pi)^{3} \delta\left(\sum_{A=1}^{3} \mathbf{p}_{A}\right) \sum_{s}a^{2}f^2 h_{\mathbf{p}_{1}}^{s} e^{i j, s}\left(\hat{p}_{1}\right) p_{2 i} p_{3 j} \delta \chi_{\mathbf{p}_{2}} \delta \chi_{\mathbf{p}_{3}}, 
\label{equ:cubicHam}
\end{aligned}
\n
\m
\begin{aligned}
H_{\mathrm{int}}^{(4)}=&-\frac{1}{4} \prod_{A=1}^{4}\left(\int \frac{\mathrm{d}^{3} p_{A}}{(2 \pi)^{3}}\right)(2 \pi)^{3} \delta\left(\sum_{A=1}^{4} \mathbf{p}_{A}\right) \sum_{s_{1}, s_{2}} a^{2}f^2 e^{i k, s_{1}}\left(\hat{p}_{1}\right) e_{k}^{j, s_{2}}\left(\hat{p}_{2}\right) p_{3 i} p_{4 j} h_{\mathbf{p}_{1}}^{s_{1}} h_{\mathbf{p}_{2}}^{s_{2}} \delta \chi_{\mathbf{p}_{3}} \delta \chi_{\mathbf{p}_{4}}.
\label{equ:quarticHam}
\end{aligned}
\n
Note that the coupling function $f(\phi)$ enters the interaction Hamiltonian as a time-dependent prefactor, making the {non-canonical spectator field} model distinct from minimally coupled cases \cite{Ota:2022xni}.
{For later use, let us make explicit the approximation adopted for the spectator perturbation.
As in \cite{Meng:2022low}, we work in the light-spectator regime in which the effective spectator mass satisfies $m/f\ll H$ during inflation, and the background motion of the spectator field is negligible. In this regime, the mixing between $\delta\chi$ and the inflaton perturbation is negligible, and the quadratic action for the spectator fluctuation reduces to
\begin{equation}
S^{(2)}_{\delta\chi}
=
\frac{1}{2}\int d\tau d^3x\, a^2 f^2
\left[
(\delta\chi')^2-(\nabla\delta\chi)^2-\frac{m^2 a^2}{f^2}\delta\chi^2
\right].
\label{eq:delta_chi_quadratic}
\end{equation}
In Fourier space, the corresponding mode equation is
\begin{equation}
\delta\chi_q''+
2\left(\frac{a'}{a}+\frac{f'}{f}\right)\delta\chi_q'
+
\left(
q^2+\frac{m^2 a^2}{f^2}
\right)\delta\chi_q
=0.
\label{eq:delta_chi_mode}
\end{equation}
Introducing the canonically normalized variable
\begin{equation}
\sigma_q \equiv a f\, \delta\chi_q ,
\label{eq:sigma_def}
\end{equation}
one obtains
\begin{equation}
\sigma_q''+
\left[
q^2-\frac{(af)''}{af}+\frac{m^2 a^2}{f^2}
\right]\sigma_q
=0.
\label{eq:sigma_mode_general}
\end{equation}
Although the mass term is kept in the action for completeness, the loop calculation in the present paper is performed in the light-spectator regime.
In this limit, the term $m^2 a^2/f^2$ is subleading compared with the effective potential $(af)''/(af)$ around the feature and horizon crossing, and the mode equation simplifies to
\begin{equation}
\sigma_q''+
\left[
q^2-\frac{(af)''}{af}
\right]\sigma_q
=0.
\label{eq:sigma_mode_light}
\end{equation}
We solve Eq.~(\ref{eq:sigma_mode_light}) numerically with the Bunch--Davies initial condition
\begin{equation}
\sigma_q(\tau)\rightarrow \frac{e^{-iq\tau}}{\sqrt{2q}},
\qquad -q\tau\gg 1.
\label{eq:sigma_BD}
\end{equation}
}

The spectator and tensor perturbations are quantized in the standard way,
\m
\begin{aligned}
\delta \chi_{\mathbf{q}}(\tau) & =u_{q}(\tau) a_{\mathbf{q}}+u_{q}^{*}(\tau) a_{-\mathbf{q}}^{\dagger},
\end{aligned}
\n
\m
\begin{aligned}
h_{\mathbf{q}}^{s}(\tau) & =v_{q}(\tau) b_{\mathbf{q}}^{s}+v_{q}^{*}(\tau) b_{-\mathbf{q}}^{s \dagger}.
\end{aligned}
\n
The creation and annihilation operators satisfy the usual commutation relations, $[a_{\mathbf{q}}, a_{-{\mathbf{q}'}}^{\dagger}]=(2 \pi)^{3} \delta(\mathbf{q}+{\mathbf{q}'})$ and $[b_{\mathbf{q}}^{s} b^{{s'} \dagger}_{-{\mathbf{q}'}}]=(2 \pi)^{3} \delta^{s {s'}} \delta(\mathbf{q}+{\mathbf{q}'})$.
In the absence of features in $f(\phi)$, the mode functions reduce to the BD vacuum,
\m
\begin{aligned}
u_{q}^{\mathrm{BD}}(\tau) & =\frac{H}{\sqrt{2 q^{3}}}(1+i q \tau) e^{-i q \tau},
\end{aligned}
\n
\m
\begin{aligned}
v_{q}^{\mathrm{BD}}(\tau) & =\frac{2 H}{M_{\mathrm{pl}} \sqrt{2 q^{3}}}(1+i q \tau) e^{-i q \tau}.
\end{aligned}
\n
Due to the {non-canonical coupling}, the scalar mode equation no longer admits an analytic solution. Therefore we compute the exact mode functions $\sigma_q(\tau)$ numerically with BD initial conditions set deep inside the horizon.

\section{Loop corrections}
\label{secIII}
\begin{figure}[!htbp]
	\centering
	\begin{minipage}[b]{0.3\textwidth}
	    \centering
	    \includegraphics[width=\textwidth]{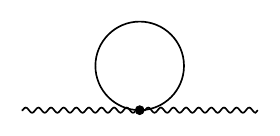}
	    \par (a) Seagull Diagram ($P_{h1}$)
        \label{fig:seagull}
    \end{minipage}
	\begin{minipage}[b]{0.3\textwidth}
	    \centering
	    \includegraphics[width=\textwidth]{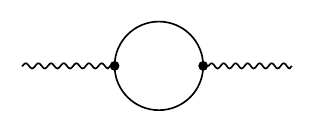}
	    \par (b) Bubble Diagram ($P_{h2}$)
        \label{fig:bubble}
    \end{minipage}
    
	\caption{One-loop Feynman diagrams correspond to $P_{h1}$ and $P_{h2}$, respectively.}
	\label{fig:feynmandiagrams}
\end{figure}

In this section, we compute the one-loop corrections to the tensor power spectrum using the in-in (Schwinger–Keldysh) formalism\cite{Maldacena:2002vr, Weinberg:2005vy}. For an operator $\mathcal{O}(\tau)$, its vacuum expectation value at time $\tau$ is given by
\m
\begin{aligned}
\langle \mathcal{O}(\tau)\rangle & =\lim_{\tau_{0} \to -\infty(1-i \epsilon)} \langle 0| \bar{T} \exp \left(i \int_{\tau_{0}}^{\tau} \mathrm{d} \tau^{\prime} H_{\mathrm{int}, I}\left(\tau^{\prime}\right)\right) \mathcal{O}_{I}(\tau) T \exp \left(-i \int_{\tau_{0}}^{\tau} \mathrm{d} \tau^{\prime \prime} H_{\mathrm{int}, I}\left(\tau^{\prime \prime}\right)\right)|0\rangle,
\label{equ:ininvev}
\end{aligned}
\n
where all fields are in the interaction picture. 
Expanding the exponentials generates the tree-level contribution and the one-loop terms arising from the cubic and quartic interactions. The resulting contributions can be organized into the seagull (quartic) and bubble (double cubic) diagrams, following the diagrammatic structure of Ref. \cite{Ota:2022xni}, as illustrated in \Fig{fig:feynmandiagrams}. The bubble and the seagull diagrams are also summarized in Ref.~\cite{Kong:2024lac} for the single-field case.

We adopt the same notation for these two contributions as Ref. \cite{Ota:2022xni} but recompute all time integrals and kernels for the {non-canonical} coupling model considered in this work. 
The tensor power spectrum is defined through the two-point function evaluated at the end of inflation,
\m
\left.\left\langle\sum_{s=+,\times} h_{\mathbf{q}}^{s}(\tau) h_{{\mathbf{q'}}}^{s}(\tau)\right\rangle\right|_{\tau=0}=(2 \pi)^{3} \delta(\mathbf{q}+{\mathbf{q'}}) P_{h}(q).
\n
so that the result up to one-loop order takes the form
\m
P_{h}(q)=P_{h 0}(q)+P_{h 1}(q)+P_{h 2}(q).
\n
where $P_{h 1}$ and $P_{h 2}$ correspond to the seagull and bubble diagrams, respectively. 
The tree-level spectrum follows from the BD tensor mode function,
\m
\begin{aligned}\label{equ:Ph0}
P_{h 0}(q)=2\left|v_{q}(0)\right|^{2}=\frac{4 H^{2}}{M_{\mathrm{pl}}^{2} q^{3}} = \frac{2\pi^2}{q^3} \cdot \frac{2H^2}{\pi^2 M_{\mathrm{pl}}^{2}}.
\end{aligned}
\n
Introducing the linear tensor-to-scalar ratio $r_0 \equiv P_{h0}/P_{\zeta}$, we can rewrite the overall amplitude factor $H^2/M_{\mathrm{pl}}^2$, using the tree-level tensor spectrum defined above, as
\begin{equation}
\frac{H^2}{M_{\mathrm{pl}}^2}
=\frac{\pi^2}{2} \cdot r_0 \cdot \frac{q^3P_{\zeta}}{2\pi^2}
\approx\left(\frac{r_0}{0.01}\right)\times10^{-10}.
\end{equation}
where we have adopted the standard CMB normalization $q^{3}P_{\zeta}/(2\pi^{2})\simeq 2\times 10^{-9}$ at the pivot scale $q_p = 0.05h/\text{Mpc}$. Note that the power of $H^2/M_{\mathrm{pl}}^2$ counts the number of loops, as shown in next equations.

The seagull contribution arises from the quartic interaction Hamiltonian $H_{\mathrm{int}}^{(4)}$ (\Eq{equ:quarticHam}) and yields
\m
\begin{aligned}
P_{h 1}(q) =& P_{h 0}\frac{H^{2}}{M_{\mathrm{pl}}^{2}}\frac{1}{3\pi^2}\mathrm{Im} \int_{\tau_{0}}^{0} \mathrm{d} \tau^{\prime} \frac{1}{q^3}(1-iq\tau^{\prime})^2\exp(2iq\tau^{\prime}) \int p_3^{4} \mathrm{d} p_3 a\left(\tau^{\prime}\right)^{2} f\left(\tau^{\prime}\right)^{2}\left|u_{p_3}\left(\tau^{\prime}\right)\right|^{2} \\
=& P_{h 0}\frac{H^{2}}{M_{\mathrm{pl}}^{2}}\frac{1}{3\pi^2}\mathrm{Im} \int_{\tau_{0}}^{0} \mathrm{d} \tau^{\prime} \frac{1}{q^3}(1-iq\tau^{\prime})^2\exp(2iq\tau^{\prime}) \int p_3^{4} \mathrm{d} p_3 \left|\sigma_{p_3}\left(\tau^{\prime}\right)\right|^{2}
\end{aligned}
\n
where $\sigma$ is the canonical variable defined by $\delta\chi=\sigma/(af)$ and is obtained numerically \cite{Meng:2022low}. 
For numerical calculation, it is convenient to introduce the dimensionless variables
\begin{align}
    x = p_*\tau=-\frac{\tau}{\tau_*},~~\tilde{p}_i = \frac{p_i}{p_*},~~\tilde{q} = \frac{q}{p_*}
    \label{equ:dimensionlessvars}
\end{align}
where $p_* \equiv a(\tau_*)H$, corresponding to the mode that exits the horizon when $\phi=\phi_*$, and $a=-1/(H\tau)$. This definition of $p_*$ is identical to the reference scale $k_*$ introduced in Ref.~\cite{Meng:2022low}. Due to a slight difference in the definition, our variable $x$ carries an extra minus sign. Then the seagull contribution can be written as
\begin{equation}
    P_{h 1}(\tilde{q}) = P_{h 0}\frac{H^{2}}{M_{\mathrm{pl}}^{2}}\frac{1}{3\pi^2}\mathrm{Im}\int_{x_{0}}^{0} \mathrm{d} x^{\prime} \frac{1}{\tilde{q}^3}(1-i\tilde{q}x^{\prime})^2\exp(2i\tilde{q}x^{\prime}) \int \tilde{p}_3^{4} \mathrm{d} \tilde{p}_3 \left|\sigma_{\tilde{p}_3}\left(x^{\prime}\right)\right|^{2}
    \label{equ:Ph1}
\end{equation}
The prefactor of Eq.~(\ref{equ:Ph1}) shows that the seagull correction is suppressed by an extra factor of $H^{2}/M_{\mathrm{pl}}^{2}$ relative to the tree-level spectrum $P_{h0}$.  
The time integral over $x'$ captures the response of the tensor mode to the effective mass shift induced by the spectator fluctuations around the feature in $f(\phi)$, while the momentum integral weights the contribution from the enhanced scalar modes through $|\sigma_{\tilde p_3}(x')|^{2}$.  
Since the integrand scales roughly as $\tilde p_3^{4}|\sigma_{\tilde p_3}|^{2}$, the seagull term is only mildly sensitive to the UV tail of the excited spectrum and typically gives a subdominant but non-negligible correction.

Expanding the in-in formalism up to second order yields two time-ordered contributions to the bubble diagram, commonly labeled $P_{h2a}$ and $P_{h2b}$. These contributions are equivalent under the symmetry of the integration measure and the kernel, and may be combined into a single expression for the bubble contribution $P_{h2}$, which we present below for clarity,
\m
\begin{aligned}\label{equ:Ph2}
P_{h 2}(\tilde{q})=&P_{h 2 a}+P_{h 2 b}\\
=&P_{h 0}\frac{H^{2}}{M_{\mathrm{pl}}^{2}}\frac{1}{\tilde{q}^3}\int_{0}^{\infty} \mathrm{d} \tilde{p}_{2} \int_{\left|\tilde{p}_{2}-\tilde{q}\right|}^{\tilde{p}_{2}+\tilde{q}} \mathrm{d} \tilde{p}_{3} \bar{w}\left(\tilde{q} ; \tilde{p}_{2}, \tilde{p}_{3}\right)\\
&\times \int_{x_{0}}^{0} d x' \int_{x_{0}}^{x'} d x''4\mathrm{Re}\{i\[\tilde{q}x'\cos (\tilde{q}x')-\sin(\tilde{q}x')\](1-i\tilde{q}x'')\mathrm{e}^{i\tilde{q}x''}\sigma_{\tilde{p}_{2}}\left(x'\right) \sigma_{\tilde{p}_{3}}\left(x'\right)\sigma_{\tilde{p}_{2}}^{*}\left(x''\right) \sigma_{\tilde{p}_{3}}^{*}\left(x''\right)\},
\end{aligned}
\n
where
\m
\bar{w}\left(\tilde{q} ; \tilde{p}_{2}, \tilde{p}_{3}\right)\equiv \frac{\tilde{p}_{2} \tilde{p}_{3}\left(\tilde{p}_{2}^{4}-2 \tilde{p}_{2}^{2}\left(\tilde{p}_{3}^{2}+\tilde{q}^{2}\right)+\left(\tilde{p}_{3}^{2}-\tilde{q}^{2}\right)^{2}\right)^{2}}{128 \pi^{2} \tilde{q}^{5}}.
\n

In the expressions for both the seagull and bubble contributions derived above, the {non-canonical} coupling function $f(\tau)$ enters explicitly.
Aside from the background dynamics themselves, the most salient difference between our results and those presented in Ref.~\cite{Ota:2022xni} is the presence of an additional factor of $f^2(\tau)$ under the time integrals, which we have absorbed into the definition of $\sigma$.

In our {non-canonical} coupling scenario, the effective mass of the spectator field depends on $f(\phi(\tau))$, and consequently the interaction Hamiltonian carries explicit $f(\tau)$ dependence (see \Eq{equ:cubicHam} and \Eq{equ:quarticHam}). After expressing the interactions in terms of the canonically normalized variable $\sigma$, this leads to an overall factor of $f^2(\tau)$ multiplying the mode functions in both the seagull and bubble diagrams.

Physically, this extra factor modifies the weight of distinct time intervals in the loop integrals according to the detailed profile of $f(\tau)$, thereby inducing quantitative differences in the one-loop tensor power spectrum relative to the minimally coupled spectator case. This structural distinction underlies the qualitative behaviors observed in \Sec{secIV} for Models G and O, in particular the resonant feature in $P_{h2}$ for Model G and the stronger enhancement in Model O.

In the superhorizon limit $\tilde{q}\equiv q/p_*\to 0$, the one-loop contributions $P_{h1}$ and $P_{h2}$ exhibit characteristic IR scalings that can be inferred from the dominant momentum dependence of their integrands.

For the seagull contribution, the integrand in Eq.~(\ref{equ:Ph1}) contains a momentum weight proportional to $\tilde{p}^4|\sigma_{\tilde{p}}(x)|^2$. In the IR regime this leads to a scaling of the form
\begin{equation}
\frac{P_{h1}}{P_{h0}}\;\sim\;\frac{H^{2}}{M_{\mathrm{pl}}^{2}}\frac{1}{3\pi^2}\int_{x_0}^0\mathrm{d}x\frac{2x^3}{3}\int \tilde{p}^{4} \mathrm{d} \tilde{p} \left|\sigma_{\tilde{p}}\left(x\right)\right|^{2},
\label{equ:Ph1IR}
\end{equation}
where $\sigma_{\tilde{p}}$ denotes the (renormalized) spectator mode function and $P_{h0}$ is the tree-level tensor spectrum. Because the momentum integral is dominated by modes around the characteristic excitation scale and the $\tilde{q}$-dependence enters only through the overall phase factor, $P_{h1}$ approaches a constant value in the $\tilde{q}\to 0$ limit, consistent with the numerical behavior shown in \Sec{secIV}.

For the bubble contribution, the integrand is weighted by $\tilde{p}^6|\sigma_{\tilde{p}}(x)|^4$, and an explicit $\tilde{q}^{-3}$ factor appears due to the phase-space measure in the convolution integral. Thus, in the IR one finds
\begin{equation}
\frac{P_{h2}}{P_{h0}}\;\sim\;\frac{H^{2}}{M_{\mathrm{pl}}^{2}}\frac{8}{45\pi^2}\int_{0}^{\infty} \mathrm{d} \tilde{p}_{2} \int_{x_{0}}^{0} \mathrm{d} x' \int_{x_{0}}^{x'} \mathrm{d} x{''} \mathrm{Re}\[-i\tilde{p}_{2}^6x{'^3}\sigma^2_{\tilde{p}_{2}}\left(x'\right) \sigma^{*2}_{\tilde{p}_{2}}\left(x{''}\right) \],
\label{equ:Ph2IR}
\end{equation}
which also saturates to an approximately $\tilde{q}$-independent plateau when $\tilde{q}\to 0$. The relative momentum weighting in the loop terms explains why the bubble contribution generally provides the dominant one-loop correction in the models under consideration.

These analytical estimates of the IR scalings agree well with the numerical results presented in \Sec{secIV}, and provide a robust explanation for the scale-invariant plateaus observed in Model~G as $\tilde{q}\to 0$.

\section{Results and Discussions}
\label{secIV}
\begin{figure}[!htbp]
	\centering
	\includegraphics[width=0.7\columnwidth]{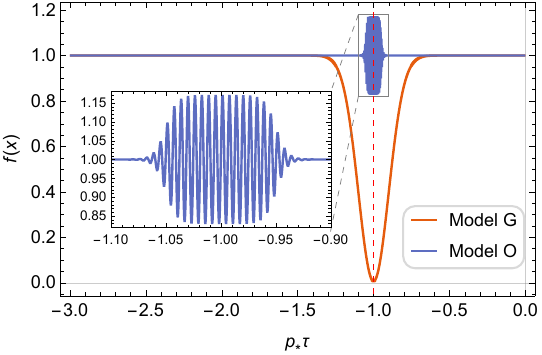}
	\caption{The {non-canonical} coupling functions $f(x)$ in \Eq{equ:modelG} and \Eq{equ:modelO}. The horizontal axis represents the dimensionless time $x=p_*\tau$.  Here we set $\Delta=0.1$, $\Lambda=0.01$, $\xi=0.001$ and the values of $A_G$ and $A_O$ are chosen for PBHs making up all of the dark matter. The vertical red dashed line marks the lower limit of the time integral, $x_0=-1$ (earlier contributions are neglected).}
	\label{fig:couplingfunctions}
\end{figure}

In this section, we present the numerical evaluation of the one-loop tensor power spectrum. We perform the time integration from $x_{0}=p_{*}\tau_{0}=-1$ to $x \to 0^{-} $ (the end of inflation). Contributions from earlier times are negligible because for $ x<x_0 $ the coupling has already set to $f \simeq 1$, and the integrand is highly oscillatory and cancels out. For numerical calculation, we write the {non-canonical} coupling functions $f(\phi)$ using the dimensionless variables \Eq{equ:dimensionlessvars} as
\begin{equation}
	f_{\mathrm{G}}(x) = 1-A_{\mathrm{G}} \exp\left[ - \frac{\left(x+1\right)^2}{2 \Delta^2} \right] 
	\label{equ:modelG}
\end{equation}
\begin{equation}
	f_{\mathrm{O}}(x)=1-\frac{A_{\mathrm{O}}}{2}\left[\tanh \frac{x+(1-\Delta / 2)}{\Lambda}-\tanh \frac{x+(1+\Delta / 2)}{\Lambda}\right] \sin \frac{x+1}{\xi}
	\label{equ:modelO}
\end{equation}
Note that there is a sign difference with Ref. \cite{Meng:2022low}, and  $\Delta\equiv\Delta_{\phi}/(\phi_*'\tau_*)$ is a constant characterizing the width of the feature in $f(\phi)$, $\Lambda\equiv\Lambda_{\phi}/(\phi_*'\tau_*)$, and $\xi\equiv\xi_{\phi}/(\phi_*'\tau_*)$. Without loss of generality, we assume $\phi_*'>0$. Similar to Ref. \cite{Meng:2022low}, we take $\Delta=0.1$, $\Lambda=0.01$, $\xi=0.001$, and $A_G$, $A_O$ are chosen such that PBHs constitute all of the dark matter,  with a mass function peaking at $10^{-12}M_{\odot}$. The coupling functions $f(x)$ for these two models are shown in \Fig{fig:couplingfunctions}. Note that Model G is characterized by a single localized ``dip'' in the coupling function and Model O involves a coupling function that oscillates rapidly.

{The parameter choices in $f(\phi)$ are not intended to span the full parameter space of the model. Instead, they are adopted from \cite{Meng:2022low}, where these parameter sets were chosen to produce a peaked primordial curvature spectrum that leads to the formation of PBHs with masses around $10^{-12}M_\odot$, capable of accounting for all the dark matter.}

{In the present work, we use these parameter sets as representative benchmark examples in which the spectator field exhibits a significant enhancement, allowing us to investigate the corresponding one-loop correction to the tensor power spectrum.}

With the {non-canonical} coupling functions specified above, we now proceed to solve for the spectator field mode functions and evaluate the one-loop contributions. To this end, we denote by $\sigma_p(\tau)$ the exact solution of the mode equation
\begin{equation}
\sigma_p''(\tau)+\Big[p^2-\frac{z''}{z}(\tau)\Big]\,\sigma_p(\tau)=0,
\qquad z(\tau)\equiv a(\tau) f(\tau).
\label{eq:sigma-eom}
\end{equation}
At high frequency $p\gg \mathcal{H}$ the mode must approach the BD vacuum,
\begin{equation}
\sigma^{\rm BD}_p(\tau)
=\frac{1}{\sqrt{2p}}\left(1-\frac{i}{p\tau}\right) e^{-ip\tau}
\label{eq:BD}
\end{equation}
Similar to \cite{Braglia:2025qrb,Braglia:2025cee}, we adopt a renormalization method to remove the UV divergence. We subtract the terms that only reproduce the BD vacuum, we introduce a momentum cutoff $\Lambda$ and define an alignment factor by matching the mode function to the BD mode function in the UV regime:
$\alpha\equiv{\sigma_{p=\Lambda}(\tau)}/{\sigma^{\rm BD}_{p=\Lambda}(\tau)}$.
The renormalized mode is then defined by
\begin{equation}
\sigma^{\rm R}_p(\tau)
\;\equiv\;
\lim_{\Lambda\to \infty}\left(\sigma_p(\tau)-\alpha \sigma^{\rm BD}_p(\tau)\right).
\label{eq:sigmaR-def}
\end{equation}
By such construction, the renormalized mode $\sigma^{\rm R}_p$ vanishes in the deep UV limit$(p\to\infty)$. This ensures that the purely BD vacuum contribution is removed, and the residual spectrum captures the physical enhancement induced by the feature in $f(\phi)$.

\begin{figure}[!htbp]
	\centering
	\includegraphics[width=1.0\columnwidth]{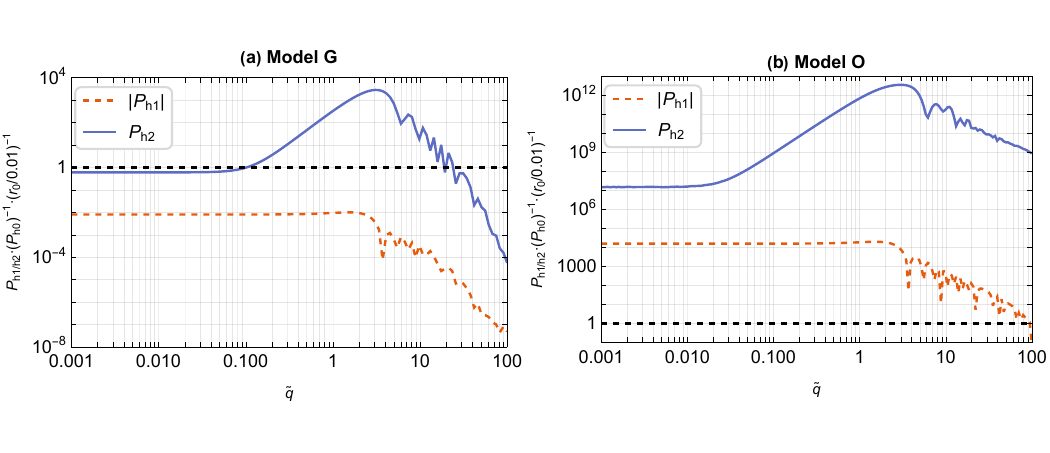}
	\caption{The one-loop corrections to the tensor power spectrum for (a) Model G and (b) Model O. The vertical axis displays the loop contributions normalized by the tree-level spectrum $P_{h0}$ and the scaling factor $(r/0.01)^{-1}$, while the horizontal axis represents the dimensionless wavenumber $\tilde{q} \equiv q/p_*$. The solid blue curves correspond to the bubble diagram contribution ($P_{h2}$), and the dashed orange curves represent absolute value of the seagull diagram contribution ($|P_{h1}|$). The horizontal black dashed line marks unity ($P_{\text{loop}} \approx P_{\text{tree}}$). Note the varying vertical scales: Model O exhibits a strong enhancement ($\sim 10^{12}$), whereas Model G shows a relative milder feature.}
	\label{fig:combfigs}
\end{figure}

 With the renormalized mode functions thus defined, we are now ready to perform the numerical integration of the loop contributions. The numerical results for both Model G and Model O are presented in \Fig{fig:combfigs}, which reveal distinct spectral features for the two scenarios.
For Model G (panel a), the loop corrections exhibit a scale-invariant plateau in the infrared regime ($\tilde{q} \ll 1$), which is consistent with analytical IR results \Eq{equ:Ph1IR} and \Eq{equ:Ph2IR}, followed by a broad resonant peak around the characteristic scale $\tilde{q} \sim \mathcal{O}(10)$. The amplitude of the corrections remains relatively moderate in the IR.
In sharp contrast, Model O (panel b) demonstrates a strong enhancement, with the one-loop power spectrum exceeding the tree-level amplitude by orders of magnitude ($\sim 10^{12}$). This clearly signals strong enhancement in the oscillatory scenario.
It is also worth noting that in both models, the bubble diagram contribution ($P_{h2}$, solid blue) consistently dominates over the seagull diagram ($|P_{h1}|$, dashed orange).

On large scales, the integrated function of the one-loop tensor power (Eq.~(\ref{equ:Ph1IR}) and Eq.~(\ref{equ:Ph2IR})) scales as $\tilde{p}^{4} \left|\sigma_{\tilde{p}}\left(x\right)\right|^{2} \sim \tilde{p}^3$ for $P_{h1}$ and $\tilde{p}^6 \left|\sigma_{\tilde{p}}\left(x\right)\right|^{4}\sim \tilde{p}^4$ for $P_{h2}$ respectively in the UV regime. This indicates the weight for higher frequency modes is stronger.
Since the characteristic momentum $p_{\rm peak}^{(O)}\gg p_{\rm peak}^{(G)}$, Model O acquires a large enhancement due to the UV scaling.
{The physical origin of the strong enhancement in Model O deserves further clarification. In the oscillatory model, the coupling function $f_O$ excites the spectator fluctuations through a resonance mechanism. As already shown in \cite{Meng:2022low}, the amplified spectator spectrum peaks around the resonance band (for our $f_O$, the peaks are around $1/(2\xi)\sim500$) rather than around the scale $\tilde q \sim 1$. Therefore, the dominant support of the loop integrals does not come from the external tensor mode crosses the feature at $\phi=\phi_\ast$. Instead, it comes from the resonantly amplified internal scalar modes.}

{A related issue is the perturbative interpretation of the large loop correction in Model O. We stress that the condition $P_{h1},P_{h2} \gtrsim P_{h0}$ does not imply a breakdown of perturbation theory. The fact that the scalar-induced one-loop contribution exceeds the tree-level vacuum fluctuation simply indicates that the primordial tensor signal in this momentum band is entirely dominated by scalar sourcing rather than by the standard inflationary vacuum modes. A genuine breakdown of perturbation theory would require the perturbative expansion within the relevant sector to become uncontrolled. For the tensor spectrum, the validity of the expansion depends on whether the two-loop contribution (sourced by the $\mathcal{O}(\delta\chi^4)$ interactions) overtakes the one-loop contribution, i.e., $P_h^{\text{2-loop}} \gtrsim P_h^{\text{1-loop}}$. However, as we have shown in previous studies \cite{Meng:2022low,Meng:2022ixx}, the enhanced spectator field fluctuations remain highly Gaussian, meaning the connected correlators are perturbatively suppressed ($\langle \delta\chi^4 \rangle_c \ll \langle \delta\chi^2 \rangle^2$). Therefore, the higher-loop corrections to the tensor spectrum remain subdominant.}

{To further check the validity of perturbation theory would require a detailed evaluation of exact second-order loop corrections and a rigorous bound on the backreaction of the amplified scalar energy density ($\rho_{\delta\chi} \ll 3 M_{\text{pl}}^2 H^2$), which is beyond the scope of the present work.}

Then we discuss our one-loop corrected primordial tensor spectrum within observations.
Based on our results up to one-loop corrections, the effective tensor-to-scalar ratio is given by
\begin{equation}
r_{\rm eff}(q)\equiv \frac{P_h(q)}{P_\zeta(q)}
= \frac{P_{h0}(q)+P_{h1}(q)+P_{h2}(q)}{P_{\zeta}(q)}.
\label{eq:reff_def}
\end{equation}
Introducing the tree-level tensor-to-scalar ratio $r_0\equiv P_{h0}/P_{\zeta}$ defined in \Sec{secIII} this leads to 
\begin{equation}
    r_{\text{eff}}(q_p) =  \frac{P_{h0}+P_{h1}+P_{h2}}{P_{h0}} \cdot \frac{P_{h0}}{P_{\zeta 
    }} = 
    r_0 \left( \frac{P_{h0}+P_{h1}+P_{h2}}{P_{h0}} \right),
\end{equation}
evaluated at the pivot scale $q_p = 0.05h/\text{Mpc}$, so that $r_{\text{eff}}(q_p)$ quantifies the effective tensor-to-scalar ratio relevant to the CMB constraints.

For Model G, the loop corrections satisfy $|P_{h1}+P_{h2}|\lesssim P_{h0}$ over the relevant scales, and thus $r_{\rm eff}(q_p)$ is only mildly shifted from $r_0$ (see Fig.~\ref{fig:combfigs} (a)). To be consistent with the CMB observations, we require $r_{\rm eff}(q_p) \lesssim 0.01$ so that
\begin{equation}
r_{\rm eff}(q_p) \lesssim 0.01 \quad \to \quad r_0 \lesssim 6.2\times 10^{-3} \qquad \text{(Model G)}.
\label{eq:reff_G}
\end{equation}
In contrast, Model O exhibits a dramatic enhancement where the one-loop corrections overwhelm the tree-level spectrum by several orders of magnitude. Taking $r_{\rm eff}$ as a reference value, we find
\begin{equation}
r_{\rm eff}(q_p) \lesssim 0.01 \quad \to \quad r_0 \lesssim 10^{-6} \qquad \text{(Model O)}.
\label{eq:reff_O}
\end{equation}

\begin{figure}[!htbp]
	\centering
	\includegraphics[width=0.7\columnwidth]{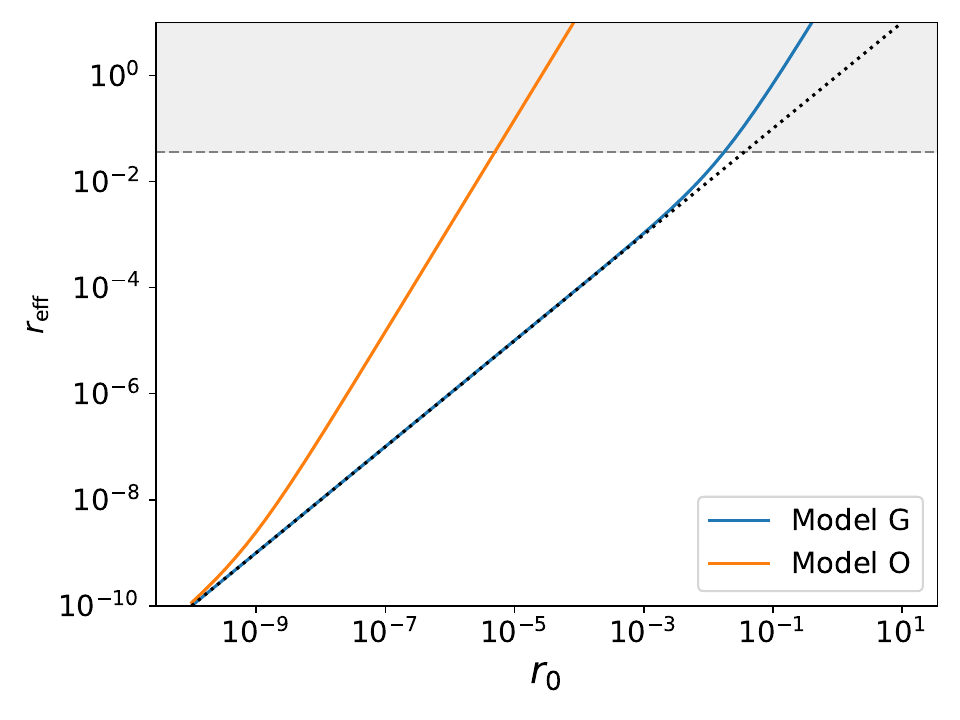}
	\caption{The effective tensor-to-scalar ratio for Model G and Model O as a function of the tree-level value. The gray shaded region denotes the $0.036$ bound given by \cite{BICEP:2021xfz} at $95\%$ C.L. and the black dotted line refers to $r_{\mathrm{eff}} = r_0$.
	}
	\label{fig:reff}
\end{figure}

The effective tensor-to-scalar ratio is demonstrated in Fig.~\ref{fig:reff}. It can be seen that Model G yields at most an $\mathcal{O}(1)$ correction to the tensor-to-scalar ratio within the observational bound while Model O leads to significant enhancement of the tensor-to-scalar ratio.

Furthermore, we compute the present-day spectral energy density of primordial GWs, $\Omega_{\text{GW}}(f)$, for modes that re-enter the horizon during radiation domination, including the effect of the relativistic degrees of freedom \cite{Watanabe:2006qe}:
\begin{equation}
    \Omega_{\text{GW,0}}\left(\eta_0, q>q_{\text{eq}}\right)=\Omega_h\left(\eta_{\text{hc}}, q\right) \Omega_{r 0}\left[\frac{g_{* s}\left(T_{\text{hc}}\right)}{g_{* s 0}}\right]^{-4 / 3}\left[\frac{g_*\left(T_{\text{hc}}\right)}{g_{* 0}}\right],
\end{equation}
where $\Omega_r$ denotes the energy density of radiation and the subscript ``0'' denotes the present-day value and ``eq'' denotes the radiation-matter equality. $\Omega_h\left(\eta_{\text{hc}}, q\right)$ is the energy density of primordial GWs for modes that re-enter the horizon during radiation domination, 
\begin{equation}
    \Omega_h\left(\eta_{\text{hc}}<\eta_{\text{eq}}, q>q_{\text{eq}}\right)=\frac{\Delta_{h, \text{prim}}^2 a^2}{12 H_{\text{eq}}^2 a_{\text{eq}}^4} q^2\left[j_1(q \eta_{\text{hc}})\right]^2.
\end{equation}
Here, $\eta_{\text{hc}} < \eta_{\text{eq}}$ is the time at horizon crossing and $T = T_{\text{hc}}$ the temperature. The dimensionless tensor power spectrum is given by $\Delta^2_{h,\text{prim}} \equiv ({q^3}/{2\pi^2}) P_{h}$ and $j_1(x)$ is spherical Bessel type function.

\begin{figure}[!htbp]
	\centering
	\includegraphics[width=0.7\columnwidth]{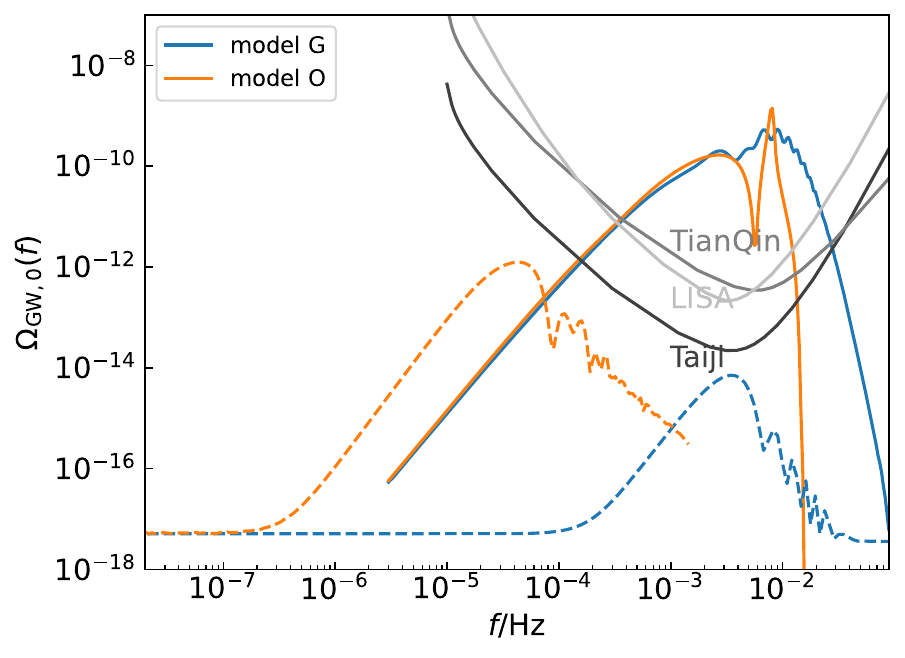}
	\caption{The present-day spectral energy density of primordial GWs as a function of frequency $f$ for Model G (blue curves) and Model O (orange curves). 
	The dashed lines are the primordial GWs and the solid lines denote scalar-induced GWs generated by Model G and Model O \cite{Meng:2022low}. The parameters for inflation are fixed at $\Delta = 0.1$, $\Lambda=0.01$, $\xi =0.001$. The characteristic scales are $p_*=7 \times 10^{11}\text{Mpc}^{-1}$, $p_*=9 \times 10^{9}\text{Mpc}^{-1}$ for Model G and Model O respectively so that the inflation can generate PBHs of $10^{-12}M_{\odot}$. The amplitude of the coupling function is chosen so that these PBHs make up all the dark matter \cite{Meng:2022low}. The $4$-year power-law integrated sensitivity curves are also shown for TianQin \cite{TianQin:2015yph}, Taiji\cite{Hu:2017mde} and LISA \cite{LISA:2017pwj}.
	}
	\label{fig:OmegaGW}
\end{figure}

Similar to \cite{Meng:2022low}, we consider PBHs of $10^{-12}M_{\odot}$ so that the characteristic scales are $p_*=7 \times 10^{11}\text{Mpc}^{-1}$, $p_*=9 \times 10^{9}\text{Mpc}^{-1}$ for Model G and Model O respectively and the amplitude of the coupling function is fixed to produce sufficient PBHs that make up all the dark matter. We take an effective tensor-to-scalar ratio $r_{\text{eff}}=0.01$ and the results are illustrated in Fig.~\ref{fig:OmegaGW}.

It can be seen that the primordial GWs are enhanced at certain scales due to the one-loop corrections from scalar perturbations. Even under the assumption that PBHs make up all the dark matter, the results of primordial GWs for both Model G and Model O are several orders of magnitude below the sensitivities of Taiji/TianQin/LISA, indicating such enhancements are not likely to be detected unless the sensitivities can be improved. 

\section{Conclusions}
\label{secV}
In this work, we have performed a first-principles calculation of the one-loop tensor power spectrum in a {non-canonical spectator field} scenario that can produce sufficient PBHs to make up all the dark matter. By explicitly computing the loop integrals associated with the seagull and bubble diagrams, we derive semi-analytical results for the one-loop tensor power spectrum. 

By considering two typical coupling functions, a single localized Gaussian dip (Model G) and an oscillating coupling (Model O), we found that the one-loop tensor spectrum can be enhanced by orders of magnitude on small scales. For Model G, the one-loop corrections in the IR limit are subdominant to the tree-level spectrum, while Model O yields a one-loop tensor power spectrum that overwhelms the tree level. We then take an effective tensor-to-scalar ratio, $r_{\text{eff}}=0.01$ as allowed by the CMB bounds and calculate the energy density for the primordial GWs.

We take the same parameters as in \cite{Meng:2022low} where the inflation generates PBHs of $10^{-12}M_{\odot}$ and make up all the dark matter. Although the scalar-induced GWs exceed the sensitivities of Taiji/TianQin/LISA, the enhanced primordial GWs are still below their sensitivities. Consequently, for the parameter choices that yield PBHs of $10^{-12}M_{\odot}$ and $r_{\text{eff}}=0.01$, the loop-enhanced primordial GW background remains well below the sensitivities of Taiji/TianQin/LISA.

{In this work, we only focus on one set of the parameters in $f(\phi)$. Here, we need to emphasize that the goal of this work is not to perform a global exploration of the parameter space. Such an analysis would require a dedicated numerical scan combined with observational constraints, which we will leave it for future work. Instead, our purpose is to demonstrate, in a controlled benchmark setup, how the presence of a nontrivial coupling function $f(\phi)$ can lead to large one-loop corrections to the primordial tensor spectrum, and to shed some light on the potential observability of this effect.}

{On the other hand, for the oscillatory benchmark, the large one-loop enhancement found here should be interpreted as evidence that resonantly amplified spectator fluctuations can make the scalar-sourced tensor signal important, rather than a breakdown of the perturbation theory. A full investigation of higher-loop contributions in this extreme case is left for future work.}

\vspace{5mm}
{\it Acknowledgments.} We gratefully acknowledge the contributions made by De-Shuang Meng in the early stages of this work. Z.L. would like to thank Atsuhisa Ota, Ying-Li Zhang and Cheng-Jun Fang for useful discussions. QGH is supported by the grants from NSFC (Grant No.~12475065, 12547110,  12447101) and the China Manned Space Program with grant no. CMS-CSST-2025-A01. C.Y. acknowledges the financial support provided under the European Union’s H2020 ERC Advanced Grant “Black holes: gravitational engines of discovery” grant agreement no. Gravitas–101052587. Views and opinions expressed are however those of the author only and do not necessarily reflect those of the European Union or the European Research Council. Neither the European Union nor the granting authority can be held responsible for them. C.Y. acknowledges support from the Villum Investigator program supported by the VILLUM Foundation (grant no. VIL37766) and the DNRF Chair program (grant no. DNRF162) by the Danish National Research Foundation. C.Y. thanks the Fundação para a Ciência e Tecnologia (FCT), Portugal, for the financial support to the Center for Astrophysics and Gravitation (CENTRA/IST/ULisboa) through grant No. UID/PRR/00099/2025 (https://doi.org/10.54499/UID/PRR/00099/2025) and grant No. UID/00099/2025 (https://doi.org/10.54499/UID/00099/2025).

\bibliography{./ref}

\end{document}